\documentclass{aa}
\makeatletter
\def\lnyoro{\mathrel{\mathpalette\gl@align<}}
\def\gnyoro{\mathrel{\mathpalette\gl@align>}}
\def\gl@align#1#2{\lower.6ex\vbox{\baselineskip\z@skip\lineskip\z@\ialign{$\m@th
#1\hfil##\hfil$\crcr#2\crcr\sim\crcr}}}
\makeatother

\begin{document}

   \thesaurus{05         
              (09.03.1;  
               09.04.1;  
               11.05.2;  
               11.09.4;  
               19.53.1)} 
   \title{Dust-to-Gas Ratio and Phase Transition of Interstellar
          Medium}

   \author{H. Hirashita\thanks{Research Fellow of the Japan Society
                               for the Promotion of Science}
          }

   \offprints{H. Hirashita}

   \institute{Department of Astronomy, Faculty of Science, Kyoto
              University, Sakyo-ku, Kyoto 606-8502, Japan             
\\
              email: hirasita@kusastro.kyoto-u.ac.jp
             }

   \date{Received March 4, 1999; accepted March 17, 1999}

   \maketitle

   \begin{abstract}
We discuss the time evolution of dust-to-gas mass ratio in the
context of multi-phase model of interstellar medium. Phase
transition of 
interstellar gas is considered to occur on a timescale of
$\sim 10^{7\mbox{--}8}$ yr, according to a nonlinear open system
model of interstellar medium. Since the phase transition 
changes the dust formation 
and destruction rates, the dust-to-gas ratio also fluctuates on
the same timescale. This explains the scatter of the 
dust-to-gas ratios of spiral galaxies quantitatively, though we
should note the large observational uncertainty.

   \end{abstract}

%

\section{Introduction}

Recent chemical evolution models of galaxies including the dust
content are successful in explaining the dust amount of nearby
galaxies (Wang \cite{wang};
Lisenfeld \& Ferrara \cite{lisenfeld}; Dwek \cite{dwek};
Hirashita \cite{hirashita}, hereafter
H99; see also Takagi, Arimoto, \& Vansevi\v{c}ius \cite{takagi}).
Supernovae (SNe) are the dominant source of the
formation of dust grains (Dwek \& Scalo \cite{dwekscalo}), and SN
shocks destroy grains (Jones et al.\ \cite{jones}; 
Borkowski \& Dwek \cite{borkowski}). Thus, in those models are
dust content connected with star formation histories.

In our previous work, H99, the dust-to-gas ratio was
expressed as a function of metallicity (see also
Lisenfeld \& Ferrara \cite{lisenfeld}), which is also related to star
formation histories. It confirmed the suggestion proposed by 
Dwek (\cite{dwek}) that 
 the accretion process onto preexisting dust grains is
efficient in spiral galaxies.
However, since the accretion is effective in cold clouds,
the global efficiency of the accretion depends on the fraction
of the gas in cold phase (Seab \cite{seab}; McKee \cite{mckee};
Draine \cite{draine}).
Thus, the efficiency varies
on a timescale of $\sim 10^{7\mbox{--}8}$ yr by the phase transition
of the ISM (Ikeuchi \cite{ikeuchi}; McKee \cite{mckee}).

In this Letter, we combine the framework of H99 with 
a theoretical work on multiphase ISM by
Ikeuchi \& Tomita (\cite{ikeuchitomita}), whose 
limit-cycle model of the ISM phase transition is applied to the
result in Tomita, Tomita, \& Sait\={o} (\cite{tomita}) by
Kamaya \& Takeuchi (\cite{kamaya}; hereafter KT97). In the
limit-cycle model, mass fraction of each phase oscillates continuously
because of  mass exchange among the components of the phases.  The
timescale of the
phase transition in the model is determined by a few parameters
intrinsic to a spiral galaxy; sweeping rate of SN shocks,
evaporation rate of the cold gas, and the cooling rate of the gas
heated by SN shocks. Actually, a static solution as well as the
limit-cycle solution for the
filling factors of the three components is
possible depending on the parameters.
However, the oscillatory behaviour (i.e., the limit-cycle model) of the
filling factors is supported observationally. Indeed, 
the observed scatter of the far-infrared-to-optical flux ratios of
spiral galaxies (Tomita, Tomita, \& Sait\={o} \cite{tomita}) is
interpreted through the limit-cycle model in KT97.
KT97 suggested that the fraction of the
gas mass (i.e., the mass filling factor) in the cold phase changes in
the range of 0.1 to 0.7 (or more) on the timescale of
$10^{7\mbox{--}8}$ yr (see also Korchagin, Ryabtsev, \& Vorobyov
\cite{korchagin}).

This Letter is organized as follows.
In \S 2, we investigate the variation of the dust-to-gas ratio 
due to the phase transition of ISM in spiral galaxies.
Finally, we discuss the result in \S 3.


\section{Grain growth in multiphase interstellar medium}

The ISM in a spiral galaxy is composed of
multiphase gas. McKee \& Ostriker (\cite{mckeeostriker}) constructed
the model of the ISM with three components  in
 a pressure equilibrium: the cold phase
($T\sim 10^2$ K and $n\sim 10$ cm$^{-3}$), the warm phase
($T\sim 10^4$ K and $n\sim 10^{-1}$ cm$^{-3}$), and the hot phase
($T\sim 10^6$ K and $n\sim 10^{-3}$ cm$^{-3}$). Since the mass of the
hot component is
negligible in a galactic disc compared with the others,
we only consider the warm and the cold gas.

Dwek (\cite{dwek}) and H99 showed that the accretion of heavy elements
onto preexisting
dust grains is the dominant process for the growth of the dust
content in spiral galaxies. Thus, we concentrate on the effect of
the phase transition on the accretion.

The timescale of the grain growth through the accretion of
heavy element, $\tau_{\rm grow}$, can be
estimated by the duration of the collisions between heavy-elements
atom and grains. According to Draine (\cite{draine}), 
$\tau_{\rm grow}\simeq 5\times 10^7$ yr in cold gas.
Here, we should note that the accretion process is more
effective in denser environments. The efficiency of the accretion
is proportional to the square of the gas density if
metallicity and dust-to-gas ratio are the same, since the
densities of both metal and dust contribute to the efficiency.
Therefore, among the three components of the ISM, we only consider
the accretion process in the cold gas, the densest component
of the ISM.

According to H99, the increase rate of dust mass by the
accretion process in a galaxy, $[dM_{\rm d}/dt]_{\rm acc}$, is
expressed as
\begin{eqnarray}
\left[\frac{dM_{\rm d}}{dt}\right]_{\rm acc}=
\frac{{\cal D}M_{\rm g}(1-f)}{\tau_{\rm acc}},\label{acc1}
\end{eqnarray}
where ${\cal D}$ is the dust-to-gas
mass ratio, 
 $M_{\rm g}$ is the total mass of ISM in the galaxy (i.e.,
$M_{\rm d}={\cal D}M_{\rm g}$), $f$
is the fraction of the metal in dust phase, and $\tau_{\rm acc}$
is the accretion timescale of heavy elements onto dust grains
(see Eq.~3 in H99). We note that the newly introduced parameter
$\tau_{\rm acc}$ is different from  $\tau_{\rm grow}$, since
$\tau_{\rm acc}$ is the accretion timescale averaged over all the
ISM phases. As commented above, the dust in the cold
gas dominantly contributes to the accretion process. Thus,
$[dM_{\rm d}/dt]_{\rm acc}$ can also be expressed in the following
way:
\begin{eqnarray}
\left[\frac{dM_{\rm d}}{dt}\right]_{\rm acc}=
\frac{{\cal D}X_{\rm cold}M_{\rm g}(1-f)}{\tau_{\rm grow}},\label{acc2}
\end{eqnarray}
where $X_{\rm cold}$ represents the mass fraction of the cold phase
to the total mass of ISM. Here, we assume that the values of
${\cal D}$ and $f$ are constant for each phase.
McKee (\cite{mckee}) showed
that the mixing of phases makes the difference in the
${\cal D}$ values among phases negligible. We also expect that
$f$ is treated as constant for all phases because of the mixing
(Tenorio-Tagle \cite{tenorio}).
Combining equations (\ref{acc1}) and (\ref{acc2}),  we finally obtain
\begin{eqnarray}
\tau_{\rm acc}=\frac{\tau_{\rm grow}}{X_{\rm cold}}.\label{acc3}
\end{eqnarray}

According to Ikeuchi (\cite{ikeuchi}) and KT97,
$X_{\rm cold}$ can vary with the range of
$0.1\lnyoro X_{\rm cold}\lnyoro 0.7$ in $10^{7\mbox{--}8}$ yr.
Therefore, from equation
(\ref{acc3}), we see that $\tau_{\rm acc}$ varies in the 
range of $1.4\tau_{\rm grow}\lnyoro\tau_{\rm accr}\lnyoro
10\tau_{\rm grow}$ on that timescale.


\section{Discussions}

We have shown in the previous section that the parameter
$\tau_{\rm acc}$, the typical timescale of accretion of heavy
elements onto dust grains, changes on a timescale 
of $10^{7\mbox{--}8}$ yr through phase transition of ISM. The range
of $\tau_{\rm acc}$
is estimated as $1.4\tau_{\rm grow}\lnyoro\tau_{\rm accr}\lnyoro
10\tau_{\rm grow}$, which is typically
$7\times 10^7~{\rm yr}\lnyoro  \tau_{\rm accr}\lnyoro
5\times 10^8~{\rm yr}$. This means that the parameter
$\beta_{\rm acc}$ (proportional to the efficiency of the accretion
of heavy elements onto preexisting dust grains), defined in H99,
changes by nearly an order of magnitude. Moreover, the timescale
of the variation is much shorter than the typical timescale of 
the gas consumption in a galactic disc ($\gnyoro 1$ Gyr; Kennicutt,
Tamblyn, \& Congdon \cite{kennicutt}). Thus, the dust-to-gas ratio
in a spiral
galaxy experiences a short-term ($\sim 10^{7\mbox{--}8}$ yr)
variation with the amplitude of an order of magnitude.

The short-term variation can be tested by examining nearby spiral
galaxies. The dust-to-gas ratios of the spiral galaxies shows
scatter around their mean
values even if the metallicity is almost the same
(Issa, MacLaren, \& Wolfendale \cite{issa}; see also H99).
According to Figure 1 in H99, the theoretical lines almost
reproduce the observed values. However,
the dust-to-gas ratios of the Galaxy and M31 differ by several
times. Both the galaxies lie in a range
of $5\lnyoro\beta_{\rm acc}\lnyoro 20$. This means that we can
explain the dust-to-gas ratios of these galaxies if 
$\beta_{\rm acc}$ changes by more than 4 times. Indeed, 
the discussion in 
\S 2 demonstrated that $\beta_{\rm acc}$ can change by more than
7 times on $\sim 10^{7\mbox{--}8}$ yr because of the phase change
of ISM. Thus, it is possible to explain the scatter of 
the dust-to-gas ratios of spiral galaxies by considering 
the phase transition.

As for dwarf galaxies,
we need another way to approach them, since
the heavy element accretion in dwarf galaxies
is much less efficient than spiral galaxies due to
their small metallicity (Hirashita \cite{hirashita99b}).
Because of their shallow gravitational potential, 
the mass outflow (e.g., Mac Low \& Ferrara \cite{mac}) can be
responsible for the dust-to-gas ratio spread, as emphasized by
Lisenfeld \& Ferrara (\cite{lisenfeld}).

We only have considered the dust formation process. However, we
should also consider dust destruction. The dominant
dust destruction occurs in the warm and hot phases
in which SN shock waves propagate (Seab \cite{seab}).
This means that
the destruction efficiency is expected to show anticorrelation
with $X_{\rm cold}$. If a galaxy is in a
higher-$X_{\rm cold}$ state, the dust destruction is more
inefficient whereas the dust growth is faster.
Thus, the variation of dust-to-gas ratio may become larger if
we take into account the dust destruction.

Finally, we should note that it is still probable that
the scatter is caused by observational uncertainty, since
the dust-to-gas ratio is not a direct observable.
However, from the discussion in \S 2, we can still propose
that the dust-to-gas ratio varies on a timescale of
$\sim 10^{7\mbox{--}8}$ yr by nearly an order of magnitude.

\begin{acknowledgements}
We would like to thank A. Ferrara, the referee, for careful reading
and useful comments that improved this Letter.
We acknowledge T. T. Takeuchi for kind helps and
invaluable discussions. 
We are grateful to S. Mineshige for continuous
encouragement. 
This work is supported by the Research Fellowship of
the Japan Society for the
Promotion of Science for Young Scientists. We made extensive use of
the NASA's Astrophysics Data System Abstract Service (ADS).
\end{acknowledgements}


\begin{thebibliography}{}

   \bibitem[1995]{borkowski} Borkowski K. J., Dwek E., 1995, ApJ 454, 254

   \bibitem[1990]{draine} Draine B. T., 1990, in: The Evolution of the
                          Interstellar Medium, ed.\
                          L. Blitz, ASP, San Francisco, p.\ 193

   \bibitem[1998]{dwek} Dwek E., 1998, ApJ 501, 643

   \bibitem[1980]{dwekscalo} Dwek E., Scalo, J. M., 1980, ApJ 239, 193

   \bibitem[1999a]{hirashita} Hirashita H., 1999a, ApJ 510, L99 (H99)

   \bibitem[1999b]{hirashita99b} Hirashita H., 1999b, ApJ  (submitted)

   \bibitem[1988]{ikeuchi} Ikeuchi S., 1988, Fund.\ Cosm.\ Phys.\
                           12, 255

   \bibitem[1983]{ikeuchitomita} Ikeuchi S., Tomita H., 1983, PASJ 35,
77

   \bibitem[1990]{issa} Issa M. R., MacLaren I., Wolfendale, A. W.,
1990, A\&A 236, 237

   \bibitem[1994]{jones} Jones A. P., Tielens A. G. G. M.,
Hollenbach D. J., McKee C. F., 1994, ApJ 433, 797

\bibitem[1997]{kamaya} Kamaya H., Takeuchi T. T., 1997, PASJ 49,
271 (KT97)

\bibitem[1994]{kennicutt} Kennicutt R. C. Jr., Tamblyn P.,
Congdon C. W., 1994, ApJ 435, 22

\bibitem[1994]{korchagin} Korchagin V. I., Ryabtsev A. D.,
Vorobyov E. I., 1994, Ap\&SS 220, 115

\bibitem[1998]{lisenfeld} Lisenfeld U., Ferrara A., 1998, ApJ 496, 145

\bibitem[1999]{mac} Mac Low, M.-M., Ferrara A., 1999, ApJ (in press)

\bibitem[1989]{mckee} McKee C. F., 1989, in: Interstellar Dust, eds.
L. J. Allamandola, A. G. G. M. Tielens, Kluwer, Dordrecht, p.\ 431

\bibitem[1977]{mckeeostriker} McKee C. F., Ostriker, J. P., 1977, ApJ
218, 148

\bibitem[1987]{seab} Seab C. G., 1987, in: Interstellar Processes, eds.\
D. J. Hollenbach, H. A. Thronson, Jr., Reidel, Dordrecht, p.\ 491

\bibitem[1999]{takagi} Takagi T., Arimoto N., Vansevi\v{c}ius, V.,
1999, ApJ (in press)

\bibitem[1996]{tenorio} Tenorio-Tagle G., 1996, AJ 111, 1641

\bibitem[1996]{tomita} Tomita A., Tomita Y., Sait\={o} M., 1996, PASJ
48, 285

\bibitem[1991]{wang} Wang B., 1991, ApJ 374, 456

\end{thebibliography}
\end{document}